\documentstyle[preprint,aps]{revtex}
\begin{document}
\draft
\title{
Spin-wave series for quantum  one-dimensional ferrimagnets}

\author{N. B. Ivanov }
\address{Institute  for Solid  State
Physics, Bulgarian Academy of Sciences,\\
Tzarigradsko
chaussee-72, 1784 Sofia, Bulgaria}

\date{\today}
\maketitle
\begin{abstract}

Second-order spin-wave expansions
are used to compute the ground-state energy
and sublattice magnetizations of the quantum one-dimensional
Heisenberg ferrimagnet with nearest-neighbor antiferromagnetic
interactions and two types of alternating
sublattice spins $S_1>S_2$.
It is found that in the extreme quantum cases
 $(S_1,S_2)=(1,1/2)$, $(3/2,1)$, and $(3/2,1/2)$,
the estimates for the  ground-state energy
and sublattice magnetizations differ less than $0.03\%$ for
the energy and  $0.2\%$ for the sublattice magnetizations
from the recently published
density matrix renormalization group numerical calculations.
The reported results strongly suggest that the quantum Heisenberg
ferrimagnetic chains  give
another example of a low-dimensional  quantum spin system
where the spin-wave approach demonstrates a surprising
efficiency.
\end{abstract}
\vspace{0.3in}
Recently, two scientific groups have
published theoretical results concerning the model of
the one-dimensional Heisenberg ferrimagnet (1DQHF)
containing two
alternating site spins ( $S_1>S_2$) per unitary cell
and nearest-neighbor
antiferromagnetic bonds\cite{pati1,kolezhuk,brehmer,swap}.
The presented linear spin-wave analysis demonstrates a
substantial reduction of the classical sublattice
spins imposed by the quantum fluctuations.
In the extreme quantum cases with $(S_1,S_2)=(1,1/2)$,
$(3/2,1)$, and $(3/2,1/2)$, the quantum reduction
of $S_2$ was shown to be about  $ 61\%$,
$ 46\%$, and $37 \%$, respectively.
The density matrix renormalization group
results\cite{swap} point towards a smaller reduction.
The cited values are respectively  reduced to $ 42\%$,
$36\%$, and $28\%$. For comparison, in the
square-lattice  Heisenberg antiferromagnet
the discussed reduction is about $39\%$. It is clear
that the linear theory overestimates the role of the zero-point
spin fluctuations.
In this respect, an open question is if the qualitative picture
based on the LSWT (and concerning,
in particular, the structure of the
elementary excitations) can reflect the real situation at all.
The purpose of the present paper is to throw some light on
the above problem through an explicit study of the large
$S$ series for the ground state
energy and sublattice magnetizations
up to second order in $1/S$.

The Hamiltonian of the 1DQHF with two spins $S_1>S_2$
per unitary cell and nearest-neighbor antiferromagnetic
couplings reads
\begin{equation}
H= \sum_{n,\delta}{\bf S_1}_{n}{\bf S_2}_{n+\delta},
\label{ham}
\end{equation}
where the  intergers $n$ number
the cells and the vector $\delta=\pm 1/2$
connects the two nearest-neighbor spins. The size of
the elementary cell and the exchgange interaction are unities.
 In what follows we frequently use the notations
 $S_1/S_2\equiv w $, $S_2\equiv S$.

We will use the Dyson-Maleev representation for the
site spin operators.
 After some standard procedures\cite{harris}, including the Fourier
 and Bogoliubov transformations and the normal ordering of the boson
 operators, the spin Hamiltonian can be recasted to the following form
\begin{eqnarray}
 H &=& H_0+V,\hspace{0.5cm}   V= c_1+V_2+V_{DM},\\
H_0&=&\left(-2wS^2+2Sc_0\right)N+2S\sum_k\left[
\omega_k^{(\alpha)}\alpha_k^{\dag}\alpha_k
+\omega_k^{(\beta)}\beta_k^{\dag}\beta_k\right], \\
V_2&=&\sum_k\left[ V_k^{(1)}\alpha_k^{\dag}\alpha_k
+V_k^{(2)}\beta_k^{\dag}\beta_k+V_k^{(3)}\alpha_k^{\dag}\beta_k^{\dag}
+V_k^{(4)}\alpha_k\beta_k \right],
\end{eqnarray}
where $c_0=-[1-(1/N)\sum_k\epsilon_k](w+1)/2$,
$c_1=-2(g_1^2+g_2^2)-2g_1g_2(w+1)w^{-1/2}$,
$g_1=-(1/2N)\sum_k \gamma_k \eta_k/\epsilon_k$,
$g_2=-(1/2)+(1/2N)\sum_k 1/\epsilon_k$,
$\epsilon_k=(1-\eta_k^2)^{1/2}$,
$\eta_k=2\gamma_kw^{1/2}/(w+1)$, $\gamma_k=\cos(k/2)$, and
$N$ is the number of cells.

$H_0$ is the  quadratic LSWT Hamiltonian.
The boson operators $\alpha_k$ and $\beta_k$
describe two types of elementary excitations with energies
\begin{equation}
E_k^{(\alpha,\beta)}=2S\omega_k^{(\alpha,\beta)}=
2S\left[\frac{w+1}{2}\epsilon_k\mp
\frac{(w-1)}{2}\right].
\end{equation}
The $\alpha$ excitations are gapless ($\omega_k^{(\alpha)}\sim
k^2$ for small $k$) and describe magnons in the sector with a total
spin $(S_1-S_2)N-1$, whereas
the $\beta$ excitations are gapful ($\omega_k^{(\beta)}=
w-1+O(k^2)$) and belong to  the sector $(S_1-S_2)N+1$.

The interaction $V$ contains three different
terms: The constant $c_1$ gives the first-order correction to
the ground state energy.
The quadratic interaction $V_2$
introduces four vertex
functions defined as follows
\begin{equation}
V_k^{(1)}=-\frac{g_1}{\epsilon_k}
\left(\frac{w+1}{\sqrt{w}}-2\gamma_k\eta_k
+\frac{w-1}{\sqrt{w}}\epsilon_k\right)
-\frac{g_2}{\epsilon_k}\left(2-
\frac{w+1}{\sqrt{w}}\gamma_k\eta_k\right)
\end{equation}
\begin{equation}
V_k^{(3)}=-\frac{g_1}{\epsilon_k}\left(2 \gamma_k
-\frac{w+1}{\sqrt{w}}\eta_k\right)-\frac{g_2}{\epsilon_k}
\left(\frac{w+1}{\sqrt{w}}\gamma_k+\frac{w-1}{\sqrt{w}}\gamma_k
\epsilon_k -2\eta_k\right)
\end{equation}
\begin{equation}
V_k^{(4)} =V_k^{(3)}+2g_2\frac{w-1}{\sqrt{w}}\gamma_k,
V_k^{(2)}=V_k^{(1)}-2g_1\frac{w-1}{\sqrt{w}}.
\end{equation}
For the model under consideration the off-diagonal terms
of $V_2$ do not vanish due to the inequality $S_1>S_2$.
Note also that $V_2$ is a non-Hermitian
operator for the same reason.

$V_{DM}$ is the Dyson-Maleev normal ordered
quartic interaction, containing
nine  vertex functions
$V^{(i)}=V^{(i)}_{12;34}$, $i=1,...,9$. We have adopted
the symmetric form used in Refs.\onlinecite{canali,ivanov} and
the convention
$(k_1,k_2,k_3,k_4)\equiv (1,2,3,4)$.

Now, let us  represent the series for the ground state energy
and the magnetization
of the first sublattice in the following form
\begin{equation}
\frac{E_0}{N}=-2wS^2+2Sc_0+c_1
+\frac{c_2}{2S}+...,
\end{equation}
\begin{equation}
m_1=wS+b_0+\frac{b_1}{2S}
+\frac{b_2}{(2S)^2}+....
\end{equation}
The coefficient $b_0=-g_2$ gives the spin reduction
in the LSWT.
The first-order
 correction for $m_1$ is related to
the off-diagonal terms in $V_2$, and a simple calculation
 gives the following result
\begin{equation}
b_1=-\frac{1}{2(w+1)}\frac{1}{N}\sum_k(V_k^{(3)}+
V_k^{(4)})\frac{\eta_k}{\epsilon_k^2}.
\end{equation}.

The coefficients $c_2$ and $b_2$ will be calculated
using the Rayleigh-Schrodinger perturbation theory.
A straightforward calculation gives the following two
contributions to $c_2=c_{21}+c_{22}$ resulting from $V_2$ and
$V_{DM}$, respectively
\begin{equation}
c_{21}=-\frac{1}{w+1}\frac{1}{N}\sum_k
\frac{V_k^{(3)}V_k^{(4)}}{\epsilon_k},\hspace{0.5cm}
c_{22}=-\frac{2}{w+1}\frac{1}{N^3}
\sum_{1-4}\delta_{12}^{34}
\frac{V^{(7)}_{12;34}V^{(8)}_{34;12}}{\epsilon_1+
\epsilon_2+\epsilon_3+\epsilon_4}.
\end{equation}.
Here $\delta_{12}^{34}$ is the Kronicker function.
The symmetric vertex functions $V^{(7)}_{12;34}$
and $V^{(8)}_{12;34}$
read
\begin{eqnarray}
V^{(7)}_{12;34} &=& {\cal U}_{1234}
\Big\{ x_1\left[ x_4(\gamma_{1-3}
-w^{-1/2}x_3\gamma_{1})-(w^{1/2}\gamma_{1-3-4}-x_3\gamma_{1-4})
\right] \nonumber \\
              &+& x_2\left[ x_4(\gamma_{2-3}
-w^{-1/2}x_3\gamma_{2})-(w^{1/2}
\gamma_{2-3-4}-x_3\gamma_{2-4})\right]\Big\},\\
V^{(8)}_{12;34} &=& {\cal U}_{1234}\Big\{ x_2\left[ (x_3
\gamma_{1-3}
-w^{-1/2}\gamma_{1})-x_4(w^{1/2}x_3\gamma_{1-3-4}-\gamma_{1-4})
\right]\nonumber \\
              &+& x_1 \left[ (x_3\gamma_{2-3}
-w^{-1/2}\gamma_{2})-x_4(w^{1/2}x_3\gamma_{2-3-4}-\gamma_{2-4})
\right] \Big\} ,
\end{eqnarray}
where ${\cal U}_{1234}=u_1u_2u_3u_4$, $u_k=(1+\epsilon_k)/2\epsilon_k$,
and $x_k=\eta_k/(1+\epsilon_k)$.

To calculate $b_2$ we introduce in $H$
a staggered magnetic field for $S_1$
spins through the operator $-h_1\sum_n {S_1}_n^z$.
Then the required second-order correction for $m_1$
can be deduced from $m^{(2)}_{1}=-(1/N)[\partial
E^{(2)}_0(h_1)/\partial h_{1}]{\mid}_{h_1=0}$,
where $E^{(2)}_0(h_1)$ is the second-order correction to $E_0$
in a finite $h_1$.
 As a matter of fact, $h_1$ reduces the structure factor to
$\eta_k\longmapsto \eta_k=2\gamma_k w^{1/2}/(1+w+h_1/2S)$, and
this is all one needs to
find all necessary derivatives.   In some cases,
it is better
to use an infinitesimal
perturbing field $h_1$ and the related perturbed ground state
$\mid 0_{h_1}\rangle=\mid 0\rangle+
[h_1/2S(w+1)]\sum_k (\eta_k/\epsilon_k^2)
\alpha_k^{\dag}\beta_k^{\dag}\mid 0\rangle+O(h_1^2)$.

Using the latter approach, one can easily find\cite{ivanov}  two of the
contributions to $b_2=b_{21}+b_{22}
+b_{23}+b_{24}$ related to the interaction $V_{DM}$
\begin{equation}
 b_{21} = -\frac{2}{(w+1)^2}\frac{1}{N^3}\sum_{1-4}
\delta_{12}^{34}\frac{V^{(7)}_{12;34}V^{(8)}_{34;12}}{(\epsilon_1+
 \epsilon_2+\epsilon_3+\epsilon_4)^2} \left(
\frac{1}{\epsilon_1}+\frac{1}{\epsilon_2}+\frac{1}{\epsilon_3}+
\frac{1}{\epsilon_4}\right),
\label{ma}
\end{equation}

\begin{equation}
 b_{22}= \frac{2w^{1/2}}{(w+1)^3}\frac{1}{N^3}
\sum_{1-4}\delta_{12}^{34}\frac{W_{12;34}}
{\epsilon_1+\epsilon_2+\epsilon_3+\epsilon_4},
\label{mb}
\end{equation}
\begin{equation}
W_{12;34}=V^{(7)}_{12;34}\left[
\frac{\gamma_2}{\epsilon^2_2}V^{(5)}_{34;12}+
\frac{\gamma_3}{\epsilon^2_3}V^{(2)}_{34;12}\right] \nonumber \\
       +\left[\frac{\gamma_1}{\epsilon^2_1}V^{(6)}_{12;34}+
\frac{\gamma_4}{\epsilon^2_4}V^{(3)}_{12;34}\right]V^{(8)}_{34;12}.
\end{equation}
The vertex functions appearing in the last equation have
the same structure as $V^{(7)}_{12;34}$ and $V^{(8)}_{12;34}$
and we will not present here their explicit expressions.

For the  other two contributions to $b_2$, which are related to the
interaction $V_2$, we use the first of the mentioned methods.
The result reads
\begin{equation}
b_{23}=-\frac{1}{(w+1)^2}\frac{1}{N}
\sum_k\frac{V_k^{(3)}V_k^{(4)}}{\epsilon_k^3},\hspace{0.5cm}
b_{24}=\frac{1}{(w+1)^2}\frac{1}{N}\sum_k\frac{1}{\epsilon_k}
\left[ U_k^{(3)}V_k^{(4)}+U_k^{(4)}V_k^{(3)}\right],
\end{equation}
where
$U_k^{(3)}=-2g_1\gamma_k/\epsilon_k-
2g_2(1-\gamma_k^2)\eta_k/\epsilon_k^3
-g_3[(w+1)w^{-1/2}\gamma_k+(w-1)w^{-1/2}
\gamma_k\epsilon_k-2\eta_k]/\epsilon_k$,
$U_k^{(4)}=U_k^{(3)}+2g_3(w-1)w^{-1/2}\gamma_k,
g_3=-(1/2N)\sum_k{\eta_k}^2/{\epsilon_k}^3$.
Notice that the equation $m_1+m_2=S_1-S_2$,
which is connected to the conservation
law for the total magnetization, is fulfilled order by order in the
spin-wave series, so that it is enough to know
only the corrections for one of the sublattice magnetizations.

The results for the series for a number of combinations $(S_1,S_2)$
are presented in the tables. Surprisingly, even in the extreme
quantum cases $(3/2,1)$ and $(1,1/2)$, the deviations from the
density matrix renormalization group results are less than
$0.033\%$ for the energy and $0.2\%$ for the
sublattice magnetizations. One can also notice that the
insrease of the ratio $w=S_1/S_2$ for
fixed $S_2\equiv S=1/2$ leads to a rapid  improvement of the series.
This tendency is expected because for large
$w$ the quasiclassical spins on
the first sublattice act as an effective field on the
$S_2=1/2$ spins.

It is instructive to compare the series for  1DQHF with those for
the square-lattice Heisenberg antiferromagnet (2DQHA)\cite{hamer}.
Let us take the series for $w=2$ and for the second sublattice
magnetization $m_2$.
\begin{eqnarray}
1DQHF:\hspace{0.5cm} \frac{E_0}{N}&=&-4S^2-0.436456\times(2S)
-0.024384+0.006518\times\frac{1}{2S}+...\\ \nonumber
2DQHA:\hspace{0.7cm} \frac{E_0}{N}&=&-4S^2-0.315895\times(2S)
-0.024948+0.000866\times\frac{1}{2S}+...\\ \nonumber
1DQHF:\hspace{0.4cm} -m_2&=&S-0.3048865+0.1212303\times\frac{1}{2S}-
0.0224602\times\frac{1}{(2S)^2}+...\\ \nonumber
2DQHA:\hspace{0.9cm} m&=&S-0.1966019+0\times\frac{1}{2S}
+0.00348\times \frac{1}{(2S)^2}+...\\ \nonumber
\end{eqnarray}
The $E_0$ spin-wave series for the compared models
have  similar structures.. The $1/S$ correction
in 1DQHF is somewhat larger, but note that the
LSWT reduction is also larger for the 1D model.
The values of the coefficients $c_1$ are very close
to each other.
As to the sublattice magnetizations, the main difference comes
from the lack of $b_1$ corrections in the 2DQHA which
is connected with the symmetry of the square lattice.
Note that up to the first order the quantum spin
reductions in the models are  approximately  one and the same.
As a whole, the  spin-wave series for the 1DQHF
demonstrate  features which are compatible with those of the 2DQHA
spin-wave series.
It is well-known that the spin-wave
results for the ground state parameters of the
$S=1/2$ square-lattice Heisenberg antiferromagnet
are close to the most precise numerical estimates\cite{sandvik}.
The reported results strongly suggest that the quantum Heisenberg
ferrimagnetic chains  give
another example of a low-dimensional  quantum spin system
where the spin-wave approach demonstrates a surprising
efficiency.

The author thanks J. Richter for useful discussions and
the staff of the Institut f\"{u}r Theoretische Physik
for hospitality. The stay in the Universit\"{a}t Magdeburg
was supported by DFG.

\begin{table}
\caption{ The coefficients of the spin-wave series
for the ground state energy per cell $\epsilon_0 =E_0/N$
of ferrimagnetic chains with two different
spins: $S_1=wS_2$, $S_2\equiv S$, $w>1$;
$\epsilon_0=-2wS^2+c_0(w)\times(2S)$
$+c_1(w)+c_2(w)\times(2S)^{-1}
+O\left[ (2S)^{-2})\right] $}.
\label{table1}
\begin{tabular}{cccccccc}
$(S_1,S_2)$&$w=S_1/S_2$&$c_0(w)$
&$c_1(w)$&$c_2(w)$&$\epsilon_0$
(SWT)&$\epsilon_0$ (DMRG)\cite{swap}&\\
\tableline
$(\frac{3}{2},1)$&1.5&-0.41403688(9)
&-0.03950436(9)&0.00874156(4)&-3.86320737&-3.86192\\
$(1,\frac{1}{2})$&$2$&-0.43645559(0)
&-0.02438442(5)&0.00651791(7)&-1.45432210&-1.45408\\
$(\frac{3}{2},\frac{1}{2})$&$3$&-0.45803557(4)
&-0.01179278(6)&0.00283359(4)&-1.96699477&-1.96724\\
$(2,\frac{1}{2})$&4&-0.46862597(4)
&-0.00691029(8)&0.00139788(3)&-2.47413839&\\
$(9,\frac{1}{2})$&18&-0.49305421(4)
&-0.00037529(8)&0.00002099(5)&-9.49340852&\\
\end{tabular}
 \end{table}
\begin{table}
\caption{ The coefficients of the spin-wave series
for the sublattice magnetization $m_1$
of ferrimagnetic chains with two different spins:
$S_1=wS_2$, $S_2\equiv S$, $w>1$;
$m_1=wS+b_0(w)$+$b_1(w)\times(2S)^{-1}+b_2(w)
\times(2S)^{-2}+O\left[ (2S)^{-3})\right] $}.
\label{table2}
\begin{tabular}{cccccccc}
$(S_1,S_2)$&$w=S_1/S_2$&$b_0(w)$
&$b_1(w)$&$b_2(w)$&$m_1$ (SWT)&$m_1$ (DMRG)\cite{swap}&\\
\tableline
$(\frac{3}{2},1)$&1.5&-0.46005842(4)
&0.22694904(9)&-0.02899689(7)&1.14616688&1.14427\\
$(1,\frac{1}{2})$&$2$&-0.30488650(5)
&0.12123033(0)&-0.02246016(3)&0.79388366&0.79248\\
$(\frac{3}{2},\frac{1}{2})$&$3$&-0.18644025(0)
&0.05316804(7)&-0.01006592(1)&1.35666188&1.35742\\
$(2,\frac{1}{2})$&4&-0.13512460(0)
&0.03000271(9)&-0.00504066(8)&1.88983745&\\
$(9,\frac{1}{2})$&18&-0.02818572(2)
&0.00152353(6)&-0.00007830(4)&8.97325951&\\
\end{tabular}
 \end{table}

\end{document}